\documentclass{pasj00}

\begin{document}
%\SetRunningHead{M. Ajiki et al.}{Overdensity of Galaxies around SDSS J0836+0054}
\Received{2006/01/11}
\Accepted{2006/03/10}

\title{New Supporting Evidence for the Overdensity of Galaxies around the
Radio-Loud Quasar SDSS J0836+0054 at $z$ =5.8}

\author{Masaru     \textsc{Ajiki},\altaffilmark{1}
        Yoshiaki   \textsc{Taniguchi},\altaffilmark{1}
        Takashi    \textsc{Murayama},\altaffilmark{1}\\
        Yasuhiro   \textsc{Shioya},\altaffilmark{1}
        Tohru      \textsc{Nagao},\altaffilmark{2,3}
        Shunji S.  \textsc{Sasaki},\altaffilmark{1}
        Yuichiro   \textsc{Hatakeyama},\altaffilmark{1}
        Taichi     \textsc{Morioka},\altaffilmark{1}\\
        Asuka      \textsc{Yokouchi},\altaffilmark{1}
	Mari I.    \textsc{Takahashi},\altaffilmark{1} and
        Osamu      \textsc{Koizumi}\altaffilmark{1}\\
}

\altaffiltext{1}{Astronomical Institute, Graduate School of Science, Tohoku University, \\
                 Aramaki, Aoba-ku, Sendai 980-8578}
\altaffiltext{2}{INAF --- Osservatorio Astrofisico di Arcetri,\\
                 Largo Enrico Fermi 5, 50125 Firenze, Italy}
\altaffiltext{3}{National Astronomical Observatory, \\
	Osawa, Mitaka, Tokyo 181-8588, Japan}


\maketitle

\begin{abstract}

Recently, Zheng et al. (2005) found evidence for an overdensity of 
galaxies around a radio-loud quasar, SDSS J0836+0054, at $z=5.8$
(a five arcmin$^2$ region). We have examined our deep optical imaging data
($B$, $V$, $r^\prime$, $i^\prime$, $z^\prime$, and {\it NB816}) taken
with the Suprime-Cam on the Subaru Telescope. The {\it NB816} ~ 
narrow-band filter  ($\lambda_{\rm c} = 815$ nm and $\Delta\lambda = 12$ nm)
is suitable for searching for Ly$\alpha$ emitters at $z\approx 5.7$.
We have found a new strong Ly$\alpha$ emitter
at $z \approx 5.7$ close to object B identified by Zheng et al.
Further, the non detection of the nine objects selected by Zheng et al. (2005)
in our $B$, $V$, and $r^{\prime}$ images provides supporting evidence that
they are high-$z$ objects.
 
\end{abstract}

\section{Introduction}

The growth of large scale structure in the early universe is one of the
important issues in the understanding of structure formation
in context of hierarchical models of galaxy formation (e.g., 
Press \& Schechter 1974; Peebles 1993).
Since the discovery of an overdensity region (or a proto cluster) of
galaxies in a deep survey field, SSA22 (Steidel et al. 2000; see also
Matsuda et al. 2005), deep surveys of such overdensity regions
have been intensively made, leading to findings of clustering of
galaxies at $z \sim$ 5 -- 6 (Shimasaku et al. 2003; 
Ouchi et al. 2005; Wang et al. 2005; Malhotra et al. 2005).

In addition to these findings, deep surveys
for sky areas surrounding high-redshift active galactic nuclei such
as radio galaxies and quasars also succeeded in finding overdensity
regions; e.g., (1) a radio galaxy, TN1338$-$1942 at $z=4.1$ (Venemans et al.
2002; Miley et al. 2004), (2) a radio galaxy, TN 0924$-$2201 at $z =5.2$
(Venemans et al. 2004; Overzier et al. 2005), (3) a quasar, SDSS J1030+0524
at $z = 6.28$ (Stiavelli et al. 2005), and (4) a radio-loud quasar,
SDSS J0836+0054 at $z = 5.8$ (Zheng et al. 2005).
Among these interesting overdensity regions, we present our deep
survey for Ly$\alpha$ emitters (hereafter LAEs) in the Zheng et al.'s
field surrounding the radio-loud quasar SDSS J0836+0054 at $z = 5.8$
because this quasar is one of our target high-$z$ quasars
(Ajiki 2006; Ajiki et al. 2006 in preparation).

As demonstrated in previous optical narrow-band imaging surveys,
a redshift range between $z = 5.6$ and 5.8 is one of favorable
windows for any ground-based telescopes because OH airglow 
emission lines are fairly weak at $\lambda \sim$ 815 nm, corresponding
to a Ly$\alpha$ redshift of $z \simeq 5.7$ (e.g., Hu et al. 1999, 2004; Rhoads \&
Malhotra 2001; Rhoads et al. 2003; see for a review Taniguchi et al. 2003b).
Using the prime focus camera, Suprime-Cam, on the 8.2m Subaru Telescope,
we have been promoting our narrow-band imaging survey for LAEs
in selected sky areas in which SDSS high-$z$ quasars at $z \sim 6$
are found (Fan et al. 2000, 2001, 2002, 2003); see for our earlier results, Ajiki
et al. (2002, 2003, 2004), Taniguchi et al. (2003a).

We adopt a flat universe with $\Omega_{\rm matter} = 0.3$,
$\Omega_{\Lambda} = 0.7$, and $H_0 = $70 km
s$^{-1}$ Mpc$^{-1}$. Throughout this paper,
magnitudes are given in the AB system.

\section{Observations and Data Reduction}

\subsection{Observations}

We have carried out very deep optical imaging in the field
 surrounding the quasar SDSS J0836$+$0054 at redshift 5.8
 (Fan et al. 2001),
 using the Suprime-Cam (Miyazaki et al. 2002)
 on the 8.2 m Subaru Telescope (Kaifu et al. 2000; Iye et al. 2004) on Mauna Kea.
The Suprime-Cam consists of ten 2k $\times$ 4k CCD chips and provides
 a very wide field of view,
 $34^\prime \times 27^\prime$ (0$\farcs$2 pixel$^{-1}$).
In this survey, 
we used broad-passband filters,
 $B$, $V$, $r^\prime$, $i^\prime$, and $z^\prime$.
We also used a narrow-passband filter,
 {\it NB816}, centered on 815 nm with a passband of
 $\Delta\lambda_{\rm FWHM} = 12$ nm; the wavelength corresponds
 to a redshift of 5.65 -- 5.75 for Ly$\alpha$ emission.
All observations were done under photometric conditions,
 and the seeing was between 0$\farcs$7 and 1$\farcs$1
 during our observing runs made in 2004
(see for details Ajiki 2006; Ajiki et al. 2006 in preparation).
The individual CCD data were reduced and combined using IMCAT by the standard process.

\subsection{Photometry}

\begin{figure*}
\FigureFile(160mm,200mm){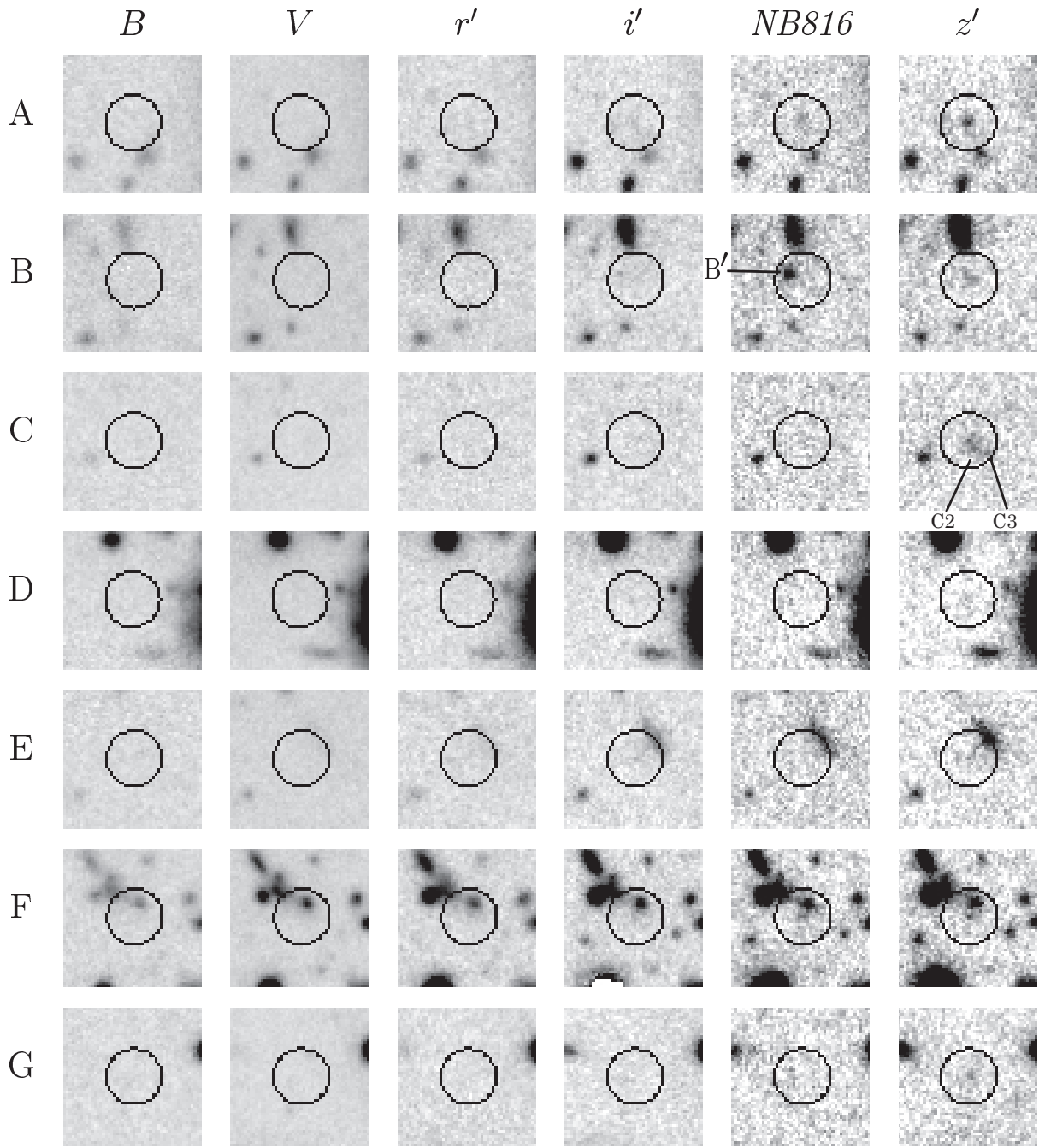}
\caption{Broad-band and {\it NB816} ~ images of the object A -- G found by Zheng et al. (2005).
         The LAE newly found in our study, B$^\prime$ is shown in
         the {\it NB816} ~ image of object B. Each box is $10''$ on a side
         (north is up and east is right).
         Each circle has $2''$ radius.\label{thum}
}
\end{figure*}

We performed photometry of the objects A -- G found by Zheng et al. (2005).
The photometry is performed with a $1^{\prime\prime}$ diameter aperture for each band
image of which image sizes of the stars have a FWHM of 1$\farcs$15.
The limiting magnitudes are 
$B=28.4$, $V=28.6$, $r^\prime=28.1$, $i^\prime=27.9$, $z^\prime=27.2$, 
and ${\it NB816}=27.3$
for a 2$\sigma$ detection with a $1^{\prime\prime}$ diameter aperture.
The reason for the use of a $1^{\prime\prime}$ diameter aperture is to
exclude possible light contamination of neighbor objects
around each target object from A to G.
Note that the light contamination in the object F could not be excluded completely
 even when we use the above small aperture.
The results of the photometry are summarized in Table \ref{tab:obj}.
Since all of objects except for the object F in Table \ref{tab:obj} are not detected 
in the $B$, $V$, and $r^{\prime}$ images above $1\sigma$ level 
(i.e.,  $B=29.1$, $V=29.3$, and $r^\prime=28.8$), we give only $i^\prime$,
$z^\prime$, and {\it NB816} ~ magnitudes in this table.
The images of the objects are shown in Figure \ref{thum}.

\section{Results and Discussion}

\subsection{A New Ly$\alpha$ Emitter at $z = 5.7$}

During the inspection of the sky area around SDSS J0836+0054,
we find a new LAE candidate close to object B (see the 2nd raw
of Figure \ref{thum}). 
It is detected in {\it NB816} ~ while not detected in the broad bands.
We refer this object as to B$^\prime$.
This object is seen at 1$\farcs$12 NE from object B.

If the excess in {\it NB816} ~ is due to Ly$\alpha$ emission,
the redshift of B$^\prime$ is $5.7\pm 0.05$ being very similar
to the photometric redshift of B, $z_{\rm ph}$(line) = 5.8 (see section 3.3).
The angular
separation (1$\farcs$12) between B and B$^\prime$ corresponds
to a projected distance of 6.5 kpc at $z=5.7$.
The angular distance of B$^\prime$ from the quasar is 74$\farcs$9.
This corresponds to a projected separation of 440 kpc.

It is likely that these two objects
are in a common dark matter halo (e.g., Hamana et al. 2004).
This system is another interacting system such as the three (or two) 
components of object C (Zheng et al. 2005) in this overdensity region.
We estimate the Ly$\alpha$ luminosity of this object,
 $L({\rm Ly}\alpha)=8.6 \times 10^{42}$ ergs s$^{-1}$,
from the {\it NB816} ~ flux in the $2''$ diameter aperture [${\it NB816}(2'')=24.8$].
This luminosity is typical among the LAEs found at $z=5.7$
 (e.g., Ajiki et al. 2003; Hu et al. 2004).

\subsection{Narrow-band photometry of Zheng et al.'s Objects}

Non detection of the nine objects selected by Zheng et al.\ (2005)
 in our $B$, $V$, and $r^{\prime}$ images provides supporting evidence
 that they are high-z objects. 
Since we have {\it NB816} ~ photometry in addition to the broad bands, we can derive precise
 information of redshifts using the narrow-band data. For instance, a LAE at $z=$5.65--5.75
 shows {\it NB816} ~ excess to the $z^{\prime}$ band 
while {\it NB816} ~ dropout objects are considered to be at $z>5.8$ (Shioya et al.~2005b).

As shown in Table 1 objects A and F were detected 2.2 $\sigma$ and 5$\sigma$ significance
 in the narrow-band image, respectively. Note that the object F is apparently affected by the foreground object seen
in the north.
 However, since one can identify their {\it NB816} ~ counterparts to the $z^\prime$ image (Fig. \ref{thum}), 
 we consider that detection of A and F is real. They are not {\it NB816} ~ dropout objects
 and thus their redshifts are probably not greater than 5.8. Since Zheng et al. (2005) estimated their
 photometric redshifts of 5.7 -- 5.8, at least the objects A and F may be associated closed to 
the quasar redshift of 5.8.

For the remaining seven objects, {\it NB816} ~ was not detected above 2 $\sigma$ level. 
However, we cannot conclude if their redshifts are grater than 5.8 or not here because the limiting
 magnitude of {\it NB816} ~ is not deep enough compared to each $z^{\prime}$ magnitude.
 Photometric redshift technique with {\it NB816} ~ and the broad bands provides possible 
redshifts of them from our photometric data (section 3.3). 

\subsection{Photometric Redshifts of Zheng et al.'s Objects}

Zheng et al. (2005) estimated photometric redshifts for the nine
objects A -- G using the Bayesian photometric redshift technique
(Ben\'itez 2000) based on their HST ACS data, $i_{775}$ and $z_{850}$. 
Since we have {\it NB816} ~ photometry for the nine objects, it is worthwhile
estimating their photometric redshifts because this trial gives an independent test
for Zheng et al.'s (2005) results. Note that our photometry is made with
a $1^{\prime\prime}$ aperture and thus the magnitudes given in Table \ref{tab:obj} are not
total magnitudes. Therefore, we use only our photometric data of $i^\prime$, $z^\prime$,
and {\it NB816}. 
 
Using the maximum likelihood described in Shioya et al. (2005a;
see for our SED templates, Nagao et al. 2004),
we estimate a most probable photometric redshift for each object. 
In this procedure, we adopt an allowed redshift of between $z=0$ and $z=7$ 
with a redshift bin of $\Delta z = 0.01$.
Our photo-$z$ analysis suggests that objects A, B, and F may be low-$z$ sources;
$z_{\rm ph} \sim 1.8$ for A, $z_{\rm ph} \sim 1.2$ for B, and $z_{\rm ph} \sim 0.24$ for F. 
These are due to that our $i^\prime - z^\prime$ colors of these three objects are
slightly bluer than those ($i_{775} - z_{850}$) of Zheng et al. (2005).
However, the photometric accuracy in Zheng et al. (2005) in their $i_{775}$ and
$z_{850}$ seems to be better than ours in $i^\prime$ and $z^\prime$.
We, therefore, adopt a $z_{\rm ph}$ whose likelihood is high at high-$z$ domain.

We do not know that the nine objects are always strong LAEs.
Therefore, we estimate two kinds of photometric redshifts; one is $z_{\rm ph}$(no line)
in which there is no Ly$\alpha$ emission, and $z_{\rm ph}$(line) in which
the effect of Ly$\alpha$ emission is taken into account.
In Table \ref{tab:obj}, we give our results together with the Bayesian photometric redshifts
($z_{\rm BP}$) obtained by Zheng et al. (2005). 

We find that our estimates are nearly
consistent with those by Zheng et al. (2005).
However, as for the three components of object C (C, C2, and C3), our results
are slightly different from theirs. This difference may be attributed to that
our photometry is affected by neighbor objects to some extent
because our image quality is not as good as HST ACS data.
Since it is likely that most galaxies at $z \sim 6$ show the Ly$\alpha$ emission
(e.g., Taniguchi et al. 2005 and references therein), we consider that 
the reliability of $z_{\rm ph}$(line) could be better than that of $z_{\rm ph}$(no line).
If this is the case, the seven objects (A, B, C3, D, E, F, and G) have 
redshifts between $z = 5.8$ and $z = 6.0$. 
Although spectroscopic follow-up observations are necessary, 
we may conclude that most of the objects found by Zheng et al. (2005)
and one LAE at $z_{\rm ph} \simeq 5.7$ found in this study
are associated with SDSS J0836+0054.

\subsection{The Overdensity of Galaxies around SDSS J0836+0054}

Based on their deep HST images, Zheng et al. (2005) found five
$i_{775}$ faint objects and two $i_{775}$ dropout in a five arcmin$^2$
sky area around SDSS J0836+0054. Comparing their finding with those
obtained with GOODS (Giavalisco et al. 2004; Dickinson et al. 2004;
Bouwens et al. 2005), they suggested that this region shows an overdensity
of galaxies by a factor of six. Although one cannot rule out the effect
of the cosmic variance, other deep surveys also find such overdensity
regions at $z \sim 6$ (e.g., Ouchi et al. 2005; Wang et al. 2005; Malhotra 
et al. 2005). Our new finding of the LAE at $z =5.7$ also
provides additional supporting evidence for the overdensity in this region.
Further investigations of probable overdensity regions in blank deep
survey fields and targeted fields around high-$z$ radio galaxies and
quasars will be important to improve our understanding of the formation
and evolution of large scale structure in the early universe.

\bigskip

We would like to thank the Subaru Telescope staff for their invaluable help.
We would also thank to the referee, Bram P. Venemans,
 for his useful comments and suggestions.
This work was financially supported in part by JSPS (Nos.
15340059, and 17253001).
MA, SSS, and TN are JSPS fellows.

%\newpage
%-------------------------------------------------------------------------

%-------------------------------------------------------------------------
%                 Table 1
%-------------------------------------------------------------------------
\begin{table*}
\begin{center}
\caption{Photometric properties of the $i$-dropout and $i$-faint objects\label{tab:obj}}
\begin{tabular}{ccccccc} 
\hline
\hline
Z05\footnotemark[$*$] &
$i^\prime$\footnotemark[$\dagger$] &
$z^\prime$\footnotemark[$\dagger$] &
${\it NB816}$\footnotemark[$\dagger$] &
$z_{\rm BP}$\footnotemark[$\ddagger$] &
$z_{\rm ph}$(no line)\footnotemark[$\S$] &
$z_{\rm ph}$(line)\footnotemark[$\|$] \\
\hline
A                    & $ 28.0^{+0.9}_{-0.5}$ & $26.4^{+0.3}_{-0.2}$ & $ 27.2^{+0.7}_{-0.4}$ & $5.8^{+1.4}_{-0.2}$ & $5.66^{+1.26}_{-0.46}$\footnotemark[$\#$] & $5.72^{+1.13}_{-0.61}$\footnotemark[$\#$] \\
B                    & $ 28.4^{+1.7}_{-0.6}$ & $27.1^{+0.6}_{-0.4}$ & $>27.3              $ & $5.9^{+1.0}_{-1.0}$ & $5.68^{+1.26}_{-0.68}$\footnotemark[$\#$] & $5.77^{+1.08}_{-0.77}$\footnotemark[$\#$] \\
C                    & $>28.1              $ & $26.6^{+0.4}_{-0.3}$ & $>27.4              $ & $5.9^{+1.1}_{-0.5}$ & $6.59^{+0.41}_{-0.91}$ & $6.73^{+0.20}_{-0.75}$ \\
C2                   & $>28.2              $ & $26.5^{+0.3}_{-0.2}$ & $>27.8              $ & $5.9^{+1.1}_{-1.5}$ & $6.14^{+0.86}_{-0.32}$ & $6.31^{+0.69}_{-0.38}$ \\
C3                   & $ 28.5^{+2.1}_{-0.7}$ & $26.9^{+0.5}_{-0.3}$ & $>28.0              $ & $7.0^{+0.0}_{-0.7}$ & $5.74^{+1.26}_{-0.06}$ & $5.78^{+1.21}_{-0.05}$ \\
D                    & $>28.5              $ & $27.8^{+2.6}_{-0.7}$ & $>28.0              $ & $5.8^{+1.2}_{-0.7}$ & $6.98^{+0.02}_{-1.27}$ & $6.01^{+0.99}_{-0.23}$ \\
E                    & $ 28.0^{+0.9}_{-0.5}$ & $27.4^{+1.0}_{-0.5}$ & $>28.0              $ & $5.2^{+1.7}_{-0.7}$ & $5.72^{+1.28}_{-0.72}$ & $5.81^{+0.21}_{-0.81}$ \\
F\footnotemark[$**$] & $ 26.9^{+0.2}_{-0.2}$ & $26.5^{+0.3}_{-0.3}$ & $ 26.3^{+0.2}_{-0.2}$ & $5.7^{+1.2}_{-0.7}$ & $5.18^{+0.10}_{-0.18}$\footnotemark[$\#$] & $5.72^{+0.05}_{-0.54}$\footnotemark[$\#$] \\
G                    & $>28.2              $ & $27.1^{+1.2}_{-0.4}$ & $>27.8              $ & $5.8^{+1.2}_{-0.8}$ & $6.96^{+0.04}_{-1.27}$ & $5.98^{+1.02}_{-0.16}$ \\ \hline
B$'$\footnotemark[$\dagger\dagger$]& $28.5^{+2.3}_{-0.7}$&  $>27.4$ & $ 25.8^{+0.1}_{-0.1}$ &                     & $5.53^{+0.06}_{-0.20}$ & $5.70^{+0.03}_{-0.05}$ \\
\hline
\multicolumn{7}{@{}l@{}}{\hbox to 0pt{\parbox{130mm}{\footnotesize
\footnotemark[$*$]{Object ID shown in Zheng et al. (2005)}
\par\noindent
\footnotemark[$\dagger$]{AB magnitude in a $1''$ diameter.
                  The total magnitude is expected to be $\gtrsim 1$ mag brighter.}
\par\noindent
\footnotemark[$\ddagger$]{Bayesian photometric redshift given by Zheng et al. (2005).}
\par\noindent
\footnotemark[$\S$]{Photometric redshift derived from the high-$z$ peak of 
                  the likelihood in which there is no Ly$\alpha$ emission.
                  The error shows 1$\sigma$ confidence level between $z=5$ and 7.
}
\par\noindent
\footnotemark[$\|$]{Photometric redshift derived from the high-$z$ peak of the likelihood 
                  in which the effect of Ly$\alpha$ emission is taken into account.
                  The error shows 1$\sigma$ confidence level between $z=5$ and 7.
}
\par\noindent
\footnotemark[$\#$]{The highest peak of the likelihood appears at $z<5$ (see text).}
\par\noindent
\footnotemark[$**$]{Photometry appears to be affected by the foreground object.}
\par\noindent
\footnotemark[$\dagger\dagger$]{LAE candidate at $z\approx5.7$ found near object B.}
}\hss}}
\end{tabular}
\end{center}
\end{table*}

\end{document}